\documentclass[sigconf,nonacm]{acmart}
\AtBeginDocument{%
  }

\usepackage{caption}
\usepackage{threeparttable}
\setcopyright{acmlicensed}
\copyrightyear{2026}
\acmYear{2026}
\acmConference[CHI EA ’26]{April 13--17,
  2026}{Barcelona}

\begin{document}

\title{Before Smelling the Video: A Two-Stage Pipeline for Interpretable Video-to-Scent Plans}

\author{Kaicheng Wang}
\affiliation{%
  \institution{University of Washington}
  \city{Seattle}
  \state{WA}
  \country{USA}
}
\email{kaichw3@uw.edu}

\author{Kevin Zhongyang Shao}
\authornote{Project co-lead.}
\affiliation{%
  \institution{University of Washington}
  \city{Seattle}
  \state{WA}
  \country{USA}
}
\email{kshao918@uw.edu}

\author{Ruiqi Chen}
\authornotemark[1]
\affiliation{%
  \institution{University of Washington}
  \city{Seattle}
  \state{WA}
  \country{USA}
}
\email{ruiqich@uw.edu}

\author{Sep Makhsous}
\affiliation{%
  \institution{University of Washington}
  \city{Seattle}
  \state{WA}
  \country{USA}
}
\email{sosper30@uw.edu}

\author{Denise Wilson}
\affiliation{%
  \institution{University of Washington}
  \city{Seattle}
  \state{WA}
  \country{USA}
}
\email{denisew@ece.uw.edu}

\renewcommand{\shortauthors}{Trovato et al.}

\begin{abstract}
 Olfactory cues can enhance immersion in interactive media, yet smell remains rare because it is difficult to author and synchronize with dynamic video. Prior olfactory interfaces rely on designer triggers and fixed event-to-odor mappings that do not scale to unconstrained content. This work examines whether semantic planning for smell is intelligible to people before physical scent delivery. We present a video-to-scent planning pipeline that separates visual semantic extraction using a vision-language model from semantic-to-olfactory inference using a large language model. Two survey studies compare system-generated scent plans with over-inclusive and naive baselines. Results show consistent preference for plans that prioritize perceptually salient cues and align scent changes with visible actions, supporting semantic planning as a foundation for future olfactory media systems.
\end{abstract}
\raggedbottom


\raggedbottom
\begin{teaserfigure}
  \centering
    \includegraphics[width=\textwidth]{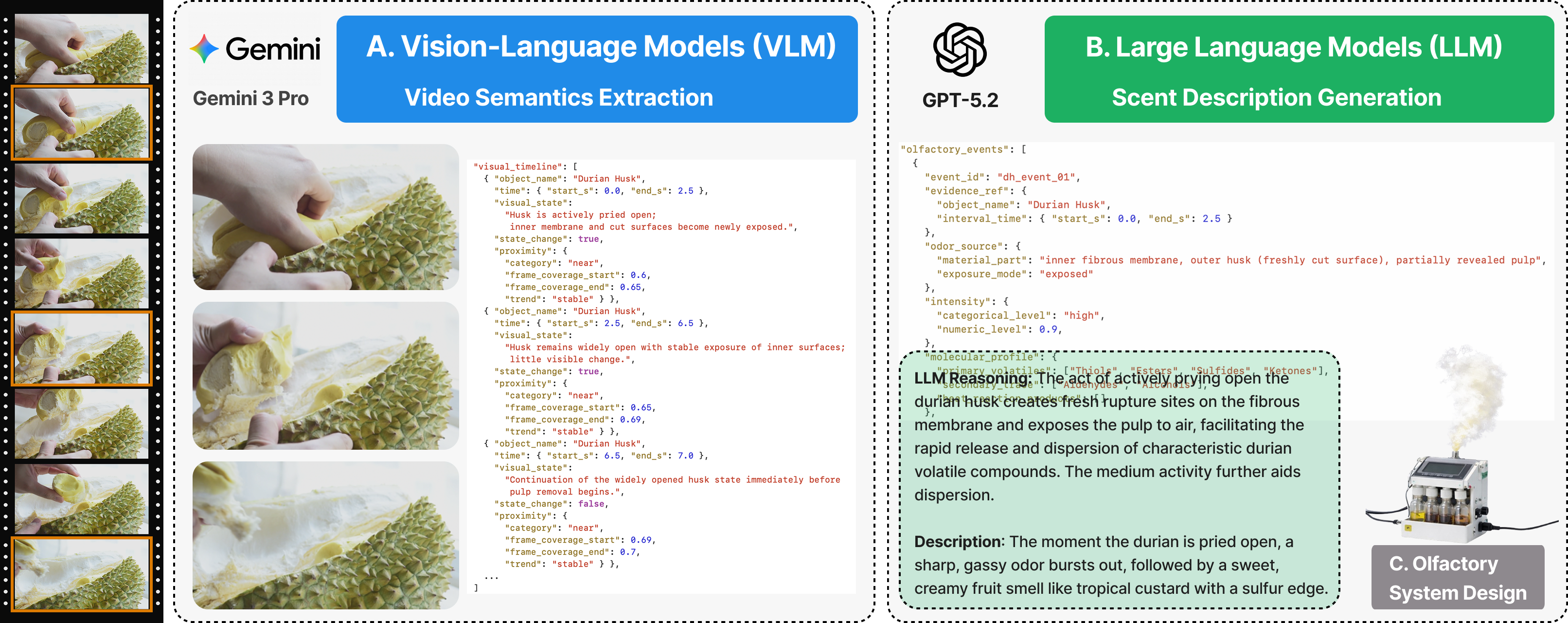}
  \caption{
  We introduce a two-stage video-to-scent planning pipeline that translates visual events in video into structured, human-interpretable scent plans, without generating physical scents. (A) A vision--language model (Gemini~3~Pro) processes uniformly sampled video frames to extract time-aligned visual semantics, producing a structured visual timeline without olfactory assumptions. (B) A large language model (GPT-5.2) transforms this visual timeline into a structured scent plan by prioritizing olfactorily relevant sources and reasoning about relative intensity and temporal evolution under a fixed odor schema. (C) The output is a temporally organized scent plan designed for future olfactory interfaces, separating semantic planning from physical scent delivery.}
  \label{fig:teaser}
\end{teaserfigure}

\maketitle

\section{Introduction}
Immersion in interactive media is shaped by how sensory channels and interaction mechanics jointly support a felt sense of presence. Video is a rich and expressive medium, yet compared to vision and sound, olfaction remains rare in deployed interactive systems, despite its strong association with emotion and memory and its demonstrated potential to increase presence in immersive media \cite{brewster2006olfoto, munyan2016olfactory, melo2020multisensory}. A core reason is that scent is difficult to integrate into dynamic media in a principled way. Scent hardware is harder to control than pixels or audio, scents linger and mix, and designers often lack reliable workflows for specifying when and what to release \cite{kaye2004making, ghinea2011olfaction, brooks2023smell}. These challenges are especially pronounced for video. Unlike static scenes, videos evolve continuously, with rapid shifts in objects, actions, and context. Because olfaction has distinct temporal dynamics, including onset, offset, and washout, “when” to present a scent can matter as much as “what” to present \cite{murray2014multiple}.


Prior olfaction-enhanced multimedia systems therefore commonly rely on manually scripted scent timelines per clip. While effective for curated demonstrations, this approach does not scale to diverse or user-generated video \cite{ghinea2010perceived, murray2014multiple}. As a result, the bottleneck is often not generating scent, but deciding which video elements are olfactorily salient and translating them into a time-varying control plan that respects human tolerance for synchronization mismatch and the practical constraints of scent delivery \cite{ghinea2011olfaction, obrist2014opportunities}. In parallel, recent vision--language models (VLMs) and video-language systems make it increasingly feasible to extract structured semantics from video, including objects, activities, and temporally grounded events \cite{krishna2017dense, wang2024internvideo2, nguyen2024video}. However, these outputs are typically optimized to describe what happened in a clip, rather than to support downstream sensory decisions. Smell-enhanced video thus requires an additional transformation layer that prioritizes a small subset of olfactorily relevant cues and converts them into a temporally organized scent plan specifying relative timing, intensity, and transitions.

In this work, we present an ongoing system exploration that investigates this semantic planning layer as a prerequisite to building end-to-end olfactory media systems in real world. We introduce a two-stage computational video-to-scent support pipeline that (1) extracts purely visual semantics from sampled video frames using VLM and (2) infers structured olfactory representations using a large language model constrained by a fixed odor schema. Rather than generating physical scents, the system produces structured scent plans intended to be intelligible, coherent, and actionable for future olfactory interfaces. To examine whether such semantic-to-scent planning aligns with human expectations, we conduct a controlled user study that evaluates system-generated scent plans without requiring physical scent output. The study compares system-generated plans against over-inclusive and naive baselines, focusing on perceived olfactory relevance, temporal coherence, and anticipated experiential impact when imagined as part of a video viewing experience. The following research questions guide our study:
\textbf{RQ1:} How well can a computational system generate scent plans that users perceive as temporally coherent and aligned with dynamic video content?
\textbf{RQ2:} Are system-generated scent plans perceived as plausible and non-disruptive when imagined as part of a video viewing experience?


\section{Related Work}
\subsection{Multisensory Augmentation and Immersion in HCI}
In Human--Computer Interaction, immersion is closely associated with presence and emerges from how sensory channels and interaction mechanics jointly support a felt sense of "being there". Prior work in immersive virtual and augmented reality shows that such systems can heighten users’ sense of presence and embodied 'being-in-place' \cite{cao2020exploratory, chen2025gestobrush, desnoyers2025being}, and that contextual factors systematically shape these experiences \cite{van2025your, jicol2023realism}. Beyond vision, auditory design, including background music, spatial audio, and adaptive sound effects, has been shown to influence emotional engagement \cite{rogers2019effects}, narrative comprehension \cite{bala2019elephant, zhao2025immersive}, and perceived realism \cite{huang2019audible, su2024sonifyar, chen2023design}.

Olfaction is strongly linked to emotion and memory, yet remains comparatively underexplored as an interaction modality. Early work such as \citet{brewster2006olfoto} highlights constraints stemming from hard-to-control hardware and inconsistent odor description schemes, while later perspectives frame smell as a pervasive but still underutilized modality in HCI, motivating olfactory interfaces for affective and cognitive experiences \cite{amores2017essence}. Recent CHI research further emphasizes the high friction involved in prototyping and iterating on olfactory experiences, which often require specialized materials, expertise, and ad hoc workflows \cite{brooks2023smell}. Despite these challenges, studies of immersive media show that even simple scent additions can increase subjective presence and immersion \cite{munyan2016olfactory, melo2020multisensory}. Compared to vision and sound, which can be tightly coupled with digital media, smell poses distinct challenges for integration \cite{brooks2023smell, kaye2004making}. In particular, olfactory cues lack a direct and reliable mapping from dynamic media content to sensory output \cite{ghinea2011olfaction, obrist2014opportunities, 10.2312:dh.20253086}. This limitation is especially pronounced in video-based scenarios, where content evolves continuously with rapid semantic transitions across scenes, objects, and activities \cite{narciso2020impact, ranasinghe2018season}. Because olfactory effects exhibit distinct temporal dynamics, the timing of scent presentation can be as consequential as scent selection itself \cite{murray2014multiple}. Without principled mechanisms for deciding when and what scent should be presented, olfactory augmentation remains difficult to scale beyond manually scripted demonstrations \cite{brooks2023smell, ranasinghe2018season}. This gap motivates closer examination of how olfactory interfaces make scent presentation decisions in dynamic, content-driven media.

\subsection{Olfactory Interfaces and Smell-Based Interaction}
Prior HCI research on olfactory interaction has largely treated scent as a discrete, designer-triggered cue. Such systems rely on designer-authored triggers to support functions such as regulating mood, conveying environmental context, or providing notifications and alerts \cite{kaye2004making, bodnar2004aroma, maggioni2018smell, dobbelstein2017inscent}. In this framing, smell is positioned as peripheral background information layered on top of audiovisual interaction \cite{kaye2004making}. While these systems establish the feasibility of olfaction as an interaction modality, their interaction logic typically assumes a small, designer-curated scent set and fixed event-to-odor mappings. Designers select scents in advance and manually specify how system states map to odors \cite{brewster2006olfoto, maggioni2020smell, brooks2023third}, resulting in sparse and predictable scent presentation with limited variation over time. This design choice is understandable given practical constraints such as hardware control, scent lingering, and user variability, but it places the semantic burden of scent selection primarily on designers rather than on the system itself \cite{ghinea2012sweet, brooks2023smell}.

This trigger-based approach becomes difficult to sustain in video-centered experiences. Video content is temporally dense and often encodes multiple overlapping semantic cues related to environment, materials, and human activity, while olfactory delivery follows different temporal dynamics including onset, offset, lingering, and washout. Prior work on olfaction-enhanced multimedia has addressed this mismatch by scripting scent events for individual clips and inserting cleanup intervals, an approach that does not translate cleanly to unconstrained or user-generated video \cite{ghinea2010perceived}. As video diversity increases, manual specification of scent timelines becomes increasingly impractical. Consequently, the core challenge for video-driven olfactory experiences shifts from physically generating scents to making semantic decisions about relevance, timing, and granularity under olfactory latency and viewers’ tolerance for synchronization mismatch \cite{maggioni2020smell, ghinea2010perceived}. This motivates computational, human-centered mechanisms that can reason over video semantics.



\subsection{From Video Semantics to Olfactory Experience}
Recent advances in video understanding through vision--language models and video LLMs make it increasingly practical to extract rich semantic information from dynamic visual content \cite{tang2025video, nguyen2024video}. These models support open-ended tasks such as question answering, summarization, retrieval, and temporally grounded event captioning \cite{krishna2017dense, wang2024internvideo2, wang2023internvid}. However, existing pipelines typically stop at recognizing and describing what happens in a clip, with outputs optimized for benchmarked objectives such as caption quality, answer accuracy, or retrieval performance \cite{tang2025video, nguyen2024video}. While effective for understanding and indexing video content, these outputs rarely provide the decisions required by olfactory systems, including what scent to emit, when to emit it, and for how long. For olfactory interaction, this level of semantic output is insufficient. Videos often contain many objects and activities, yet only a subset meaningfully contributes to olfactory perception. For example, a kitchen scene includes numerous entities, but only a few plausibly define the smellscape. Moreover, olfactory experience depends not only on the presence of elements, but also on their relative importance, temporal ordering, intensity, and duration \cite{ghinea2011olfaction, murray2014multiple, tewell2024review}. Smell generation therefore requires semantic prioritization and transformation oriented toward human perception rather than visual completeness \cite{obrist2014opportunities, maggioni2020smell}. Existing video understanding systems rarely address this experiential layer, as they do not distinguish perceptually salient cues for olfaction or translate semantics into representations suitable for downstream sensory output \cite{tang2025video, murray2014multiple}. As a result, the applicability of current video semantics pipelines for dynamic, smell-enhanced media remains limited.

\section{Methods} 
To provide early empirical grounding for our video-to-olfactory planning approach, we conducted two online survey studies. As this work reports an ongoing system prototype submitted as a poster, the studies were designed to assess the feasibility and experiential plausibility of semantic scent planning rather than to validate an end-to-end olfactory system. Together, the studies examine whether the system’s semantic decision-making aligns with human intuition and whether the resulting scent plans are perceived as reasonable and non-disruptive when imagined in use. Study~1 evaluates the system’s ability to prioritize olfactorily relevant semantics by contrasting alternative planning strategies, while Study~2 serves as a downstream plausibility check on perceived immersion and distraction.

\subsection{Survey recruitment and respondents}
Participants were recruited through exploratory sampling strategies appropriate for early-stage research. Recruitment messages were shared via Slack-based academic and student communities, and participants were encouraged to forward the study to peers through snowball sampling. All respondents were at least 18 years old and provided informed consent prior to beginning the survey. The study protocol was reviewed and approved by the ethics review board of [Anonymous University]. A total of 22 responses were collected. 8 incomplete submissions were excluded prior to analysis. Of the remaining responses, 14 participants completed Study~1 and 8 participants completed Study~2. Participants were between 18 and 24 years old. Most respondents identified as Asian (\(n=10\), 71.4\%), followed by White (\(n=4\), 28.5\%) and Middle Eastern (\(n=1\), 7.1\%).

\subsection{Survey stimuli and materials}
Both studies used a shared set of system outputs generated from short video clips. We prepared ten brief clips depicting visually and semantically rich scenes with plausible olfactory relevance. For each clip, we generated three scent plans: \textbf{(1)} a system-generated plan produced by the proposed pipeline, \textbf{(2)} an over-inclusive baseline that incorporates detected semantic elements without prioritization, and \textbf{(3)} a naive baseline based on a simplified semantic mapping strategy. All plans were presented in a standardized textual format describing the intended scent and its temporal structure. Plan order was randomized across questions to reduce bias.

\subsubsection{Study~1: Semantic selection for olfactory planning}
Study~1 examines whether the proposed system selects semantically appropriate elements to drive olfactory experience design. Participants completed ranking-based comparisons and optional open-ended prompts to capture intuitive judgments and reasoning. For each of the ten video clips, participants watched the clip and then ranked the three corresponding scent plans from most to least suitable for supporting olfactory immersion. This comparative design emphasizes relative judgments between alternative semantic selection strategies rather than absolute ratings. After each ranking, participants explained their top choice in an open-ended response, describing which moments or visual cues should influence the scent and which should not. These responses were used to identify perceived olfactory relevance, common failure modes, and points of alignment or mismatch between system decisions and human expectations.

\subsubsection{Study~2: Experiential plausibility of system-generated plans}
Study~2 provides a lightweight plausibility check on the experiential implications of system-generated scent plans. Rather than re-evaluating semantic selection, it examines whether participants perceive the plans as coherent and non-disruptive when imagined in use. Three representative video clips were selected for deeper evaluation. For each clip, participants compared two scent plans: the system-generated plan and the over-inclusive baseline. Participants were instructed to imagine the scent plans presented in synchrony with the video, following common practices in early-stage multisensory HCI research. They rated perceived immersion enhancement, coherence with the video’s progression, and perceived distraction using seven-point Likert scales, and indicated which plan they would prefer to experience. Optional open-ended questions captured reasoning about factors such as objects, actions, environment, and overall atmosphere, providing qualitative insight into perceived plausibility and potential risks of video-driven olfactory augmentation.



\section{Results and Discussion}

\textbf{Study 1: Semantic selection for olfactory planning.}
Aggregated rankings revealed a clear and consistent preference for system-generated scent plans over both baselines (Table~\ref{tab:study1_rank_desc}). The system condition achieved the lowest mean rank (1.586), followed by the over-inclusive baseline (1.871) and the naive baseline (2.543). This ordering was also reflected in first-place selections, with system-generated plans ranked first in 54.3\% of trials, compared to 32.1\% for the over-inclusive baseline and 13.6\% for the naive baseline, indicating a substantial preference margin rather than a marginal effect.

\begin{table}[ht]
  \centering
  \caption{Study~1 aggregated ranking descriptive (\(n=14\)).}
  \label{tab:study1_rank_desc}
  \small
  \setlength{\tabcolsep}{5pt}
  \begin{tabular}{lrrrrrr}
    \toprule
    Condition & Mean Rank & CI$_{low}$ & CI$_{high}$ & Rank~1 Rate & CI$_{low}$ & CI$_{high}$ \\
    \midrule
    System & 1.586 & 1.464 & 1.714 & 0.543 & 0.457 & 0.636 \\
    Over   & 1.871 & 1.736 & 2.007 & 0.321 & 0.236 & 0.407 \\
    Naive  & 2.543 & 2.378 & 2.714 & 0.136 & 0.071 & 0.207 \\
    \bottomrule
  \end{tabular}
\end{table}

A Friedman test confirmed a significant difference in aggregated ranks across conditions ($\chi^2 = 19.36$, $p = 6.26 \times 10^{-5}$). Pairwise Wilcoxon signed-rank tests with Holm correction (Table~\ref{tab:study1_rank_tests}) showed that the system condition significantly outperformed both baselines, and that the over-inclusive baseline consistently outperformed the naive baseline. Together, these results indicate a stable ordering among semantic-to-scent strategies rather than isolated pairwise differences.

\begin{table}[ht]
  \centering
  \caption{Study~1 Pairwise Wilcoxon signed-rank tests with Holm correction.}
  \label{tab:study1_rank_tests}
  \small
  \begin{tabular}{lrrr}
    \toprule
    Pair & $W$ & $p$ & $p_{\text{Holm}}$ \\
    \midrule
    System vs.\ Over  & 13.5 & 0.0251 & 0.0251 \\
    System vs.\ Naive & 0.0  & 0.000122 & 0.000366 \\
    Over vs.\ Naive   & 1.0  & 0.00287 & 0.00575 \\
    \bottomrule
  \end{tabular}
\end{table}

Qualitative responses help explain this ordering. Across videos and participants, respondents consistently prioritized a single dominant olfactory source over exhaustive coverage of all visible elements. Focusing on the primary source, such as food, flowers, or a central object, was described as making the scent easier to imagine, whereas secondary or background elements were often viewed as distracting or unnecessary. Participants also drew a clear distinction between visual presence and olfactory relevance. Visually salient elements such as cutting boards, metal surfaces, stone, air, lighting, gloves, or containers were frequently rejected as scent contributors, despite being clearly visible in the video. These elements were described as implausible to smell or as diluting the primary olfactory impression, suggesting that olfactory relevance is governed by perceptual expectations rather than visual salience alone. Participants further grounded olfactory judgments in moments of change, such as cutting, squeezing, spraying, opening, or adding ingredients, treating smell as something that emerges or intensifies over time. Concise descriptions that selectively abstracted these transitions were preferred over diffuse or overly detailed plans.

\textbf{Study 2: Experiential plausibility of system-generated plans.}
Study~2 examined whether system-generated scent plans were perceived as plausible and non-disruptive when imagined as part of a real video viewing experience. As summarized in Table~\ref{tab:study2_results} in Appendix, participants  preferred the system-generated plans over the over-inclusive plans, reporting higher immersion, lower distraction, and stronger coherence with the video's progression. Participants reported that the timing and evolution of the scent plans felt appropriate. Responses described the onset and changes of scent as “right,” “felt appropriate,” or well aligned with observed actions such as cutting, blooming, or handling objects.

When participants raised concerns about potential distraction or discomfort, these concerns were primarily attributed to specific descriptive choices rather than to the presence of olfactory augmentation itself. Overly intense terms, semantically implausible cues, or references to secondary elements such as tools or background materials were described as potentially distracting or uncomfortable if physically presented. In contrast, participants rarely questioned the idea of adding scent in principle, instead framing issues as design choices that could be adjusted.

Participants also articulated clear expectations regarding when scent would enhance immersion and when it might be unnecessary or undesirable. Scent was viewed as most appropriate for videos centered on food, nature, or atmospheric environments, and less appropriate for action-dominant or socially focused content. Across responses, participants emphasized the importance of user control, expressing a desire for explicit mechanisms to adjust intensity, timing, or to toggle scent on and off altogether. These observations suggest that the proposed planning approach is perceived as intelligible and plausible, while highlighting key sensitivities and control requirements for future system development.



\section{Conclusion and Future Work}
This work investigates a foundational prerequisite for olfactory media systems: whether semantic plans for smell derived from video content are intelligible and evaluable by people before physical scent delivery is introduced. We presented an ongoing two-stage video-to-scent planning pipeline and conducted two survey studies to examine how users interpret system-generated scent plans in terms of relevance, temporal coherence, and experiential plausibility. Results show that participants exhibit stable expectations about olfactory salience, favoring plans that prioritize dominant scent sources, align scent changes with observable actions, and avoid unnecessary or implausible descriptors. As a poster, this work focuses on validating the semantic and experiential intelligibility of scent planning rather than demonstrating end-to-end olfactory output. Future work will integrate these plans with physical scent-delivery devices, explore adaptive user controls for intensity and timing, and extend the approach to interactive and user-generated media.

\bibliographystyle{ACM-Reference-Format}
\bibliography{main}
\newpage
\appendix
\section{Appendix}\label{sec:appendix}

\begin{table}[ht]
  \centering
  \caption{Study 2 aggregated results comparing the system-generated scent plan against the over-inclusive baseline (\(n=8\)).}
  \begin{threeparttable}[ht]
  \small
  \begin{tabular}{lrrrrrrrrrr}
    \toprule
    Construct & $\Delta$Mean\tnote{1} & $\Delta$Mean CI$_{low}$\tnote{4} & $\Delta$Mean CI$_{high}$ & Preference Rate\tnote{5} & CI$_{low}$ & CI$_{high}$ & $W$\tnote{5} & $p$\tnote{6} & $p_{\text{Holm}}$\tnote{7} \\
    \midrule
    Immersion\tnote{2}   & 1.42 & 1.04 & 1.79 & 1.00 & 1.00 & 1.00 & 0.0 & 0.0078 & 0.0313 \\
    Distraction\tnote{3} & 2.17 & 1.71 & 2.83 & 1.00 & 1.00 & 1.00 & 0.0 & 0.0078 & 0.0313 \\
    Coherence\tnote{2}   & 1.29 & 0.62 & 1.83 & 0.88 & 0.62 & 1.00 & 1.0 & 0.0156 & 0.0313 \\
    Easy to imagine\tnote{2}       & 1.50 & 0.92 & 2.08 & 0.88 & 0.62 & 1.00 & 0.0 & 0.0176 & 0.0313 \\
    \bottomrule
  \end{tabular}
  \begin{tablenotes}
     \item[1] $\Delta$Mean is the participant-aggregated difference between the system and baseline plans; $\Delta>0$ indicates an advantage for the system.
     \item[2] For immersion, coherence, and easy-to-imagine, $\Delta=\text{System}-\text{Baseline}$.
     \item[3] For distraction, $\Delta=\text{Baseline}-\text{System}$ so that larger $\Delta>0$ indicates the system is perceived as less distracting.
     \item[4] CIs are 95\% bootstrap intervals.
     \item[5] Preference Rate is the proportion of participants with aggregated $\Delta>0$.
     \item[6] $W$ and $p$ are from Wilcoxon signed-rank tests.
     \item[7] $p_{\text{Holm}}$ applies Holm correction across constructs.
   \end{tablenotes}
    \end{threeparttable}%
    \label{tab:study2_results}
\end{table}

\end{document}